\IfFileExists{colortbl.sty}{\PassOptionsToPackage{table}{xcolor}}{}
\documentclass{amii}

\setcitestyle{numbers,square,comma}
\usepackage{xurl}
\usepackage{algorithm}
\usepackage{algpseudocode}
\usepackage{tabularx}
\usepackage{textcomp}
\usepackage{listings}
\usepackage{enumitem}
\usepackage{epigraph}

\setlist[itemize]{leftmargin=10pt}

\lstdefinestyle{codestyle}{
    basicstyle=\ttfamily\footnotesize,
    breaklines=true,
    showstringspaces=false,
    tabsize=4,
}

\newcommand{\ourmethod}{\textsc{RucTangle}} %
\newcommand{\tangleeval}{\textsc{TangleEval}} %

\definecolor{mdlA}{HTML}{F4F4F4}   %
\definecolor{mdlB}{HTML}{EAF0F8}   %
\definecolor{mdlC}{HTML}{F9F1E7}   %
\makeatletter
\@ifpackageloaded{colortbl}{%
  \newcommand{\rmswatch}[1]{\fcolorbox{black!35}{#1}{\rule{0pt}{4pt}\rule{7pt}{0pt}}}%
}{%
  \newcommand{\rmswatch}[1]{}%
}
\makeatother

\definecolor{keyboxframe}{RGB}{38,70,110}
\definecolor{keyboxback}{RGB}{241,245,249}

\newsavebox{\keyboxbuffer}
\newlength{\keyboxinner}

\newenvironment{keybox}[1]
  {%
    \par\medskip\noindent
    \setlength{\keyboxinner}{%
      \dimexpr\linewidth-2\fboxsep-14.5pt\relax
    }%
    \begin{lrbox}{\keyboxbuffer}%
    \begin{minipage}{\keyboxinner}%
      \normalfont\normalsize
      \vspace{3pt}%
      \noindent\textbf{#1:}\enspace\ignorespaces
  }
  {%
      \par\vspace{4pt}%
    \end{minipage}%
    \end{lrbox}%
    \noindent\colorbox{keyboxback}{%
      {\color{keyboxframe}%
        \rule[-\dp\keyboxbuffer]{2.5pt}{%
          \dimexpr\ht\keyboxbuffer+\dp\keyboxbuffer\relax
        }%
      }%
      \hspace{6pt}%
      \usebox{\keyboxbuffer}%
      \hspace{6pt}%
    }%
    \par\medskip
  }

\patchcmd{\title}{\huge}{\LARGE}{}{\PackageError{amii-preprint}{Could not set title size}{Check the Amii class title definition.}}
\title{Runnable Commit Untangling for Coding Agents}
\author[1,4]{Jinfeng Jiang}
\author[1,\dagger]{Dongsun Kim}
\author[2]{Dayi Lin}
\author[3,4,5,\dagger]{Zhou Yang}

\affiliation[1]{Korea University}
\affiliation[2]{University of Waterloo}
\affiliation[3]{University of Alberta}
\affiliation[4]{\mbox{Alberta Machine Intelligence Institute (Amii)}}
\affiliation[5]{Canada CIFAR AI Chair}

\contribution[\dagger]{Corresponding authors}
\shorttitle{Runnable Commit Untangling for Coding Agents}
\shortauthors{Jiang et al.}

\abstract{Coding agents produce \emph{large, tangled patches} that mix multiple development purposes, making the code hard to review and maintain. 
Commit untangling offers the promise of organizing such large patches into \textit{untangled, manageable commits}. 
This paper emphasizes two important limitations in existing commit untangling studies. 
First, they do not consider that untangled commits are ordered and should leave the code \emph{runnable}. In practice, maintainers are unlikely to accept commits that prevent the code from running.
Second, existing studies claim that commit untangling helps software maintenance. However, they conduct syntactic comparisons between the untangled commits and developers' original commits without directly showing the claimed maintenance benefits.
To address these gaps, this paper makes two novel contributions: \textcircled{\scriptsize 1} \ourmethod{}, the first agentic method that untangles commits while keeping the code runnable after each commit; and \textcircled{\scriptsize 2}~\tangleeval{}, the first evaluation framework that quantifies how untangled, manageable commit histories help coding agents repair bugs.
We compare \ourmethod{} against four untangling methods on 131 agent-generated patches. All histories produced by \ourmethod{} are runnable, while baselines produce 20.6\%--37.4\% unrunnable commit histories. 
We further collect 453 agent-generated patches that introduce regressions (i.e., causing previously passing tests to fail) and ask two other coding agents to repair regressions. Augmenting agent context with \ourmethod{}-produced histories yields 5.2\% absolute improvement in pass@1.
We also analyze agent trajectories to learn how they use untangled commits to navigate and fix bugs. 
Our findings demonstrate the value of adopting established software engineering practices in the era of coding agents, which broaden the future research agenda: how can agents actively use software history to make better development decisions?}

\date{October 8, 2026}
\correspondence{\email{darkrsw@korea.ac.kr}; \email{zy25@ualberta.ca}}

\begin{document}
\maketitle

\section{Introduction}
\label{sec:intro}

\begingroup

\setlength{\epigraphwidth}{0.60\linewidth}
\setlength{\epigraphrule}{0pt}
\setlength{\beforeepigraphskip}{0pt}
\setlength{\afterepigraphskip}{0.35\baselineskip}

\renewcommand{\epigraphsize}{\small}
\renewcommand{\epigraphflush}{flushright}
\renewcommand{\textflush}{flushleft}
\renewcommand{\sourceflush}{flushright}

\begin{epigraphs}
  \setlength{\itemsep}{0.4\baselineskip}

  \qitem
    {``Change the programs in small steps,
      so if you make a mistake, it is easy to find where the bug is.''}
    {---\,\textit{Martin Fowler}%
      ~\cite{fowler2018refactoring}}

\end{epigraphs}

\endgroup

Coding agents are increasingly used in software development, but their output still requires review and maintenance~\cite{wang2026rejected}. 
Google reported that AI generated 75\% of its new code in April 2026, with engineers reviewing and approving the output~\cite{pichai2026}.
Agent-written patches are often \emph{tangled}, containing changes that serve multiple development purposes, such as implementing a feature and refactoring existing code~\cite{tian2026refactoringrunaway,horikawa2025agenticrefactoring}. 
These patches are also substantially larger: on SWE-bench Verified, agent-written patches are twice as large as human-written patches~\cite{luo2026recap}.
When different purposes are mixed in a large patch, reviewers must distinguish the intended behavior changes from edits that should preserve existing behavior.
If the patch introduces a bug, developers must also identify which of these changes caused it.
Organizing these changes so that they can be inspected separately is therefore important for reviewing and maintaining agent-generated code.

\emph{Commit untangling} aims to organize a tangled patch into smaller commits, each containing changes that serve a single development purpose~\cite{shen2021smartcommit,partachi2020flexeme,li2022utango}.
Prior work argues that separating changes by purpose makes them easier to understand and review~\cite{shen2021smartcommit,li2022utango}.
These studies also motivate untangling through its potential benefits for software maintenance.
For example, separating unrelated changes can help developers narrow down which changes introduced a bug~\cite{partachi2020flexeme}.
It can also let developers revert an unwanted feature without discarding an unrelated bug fix included in the same patch~\cite{shen2021smartcommit,atomizer2026}.
However, this paper emphasizes \textit{two key limitations in how existing studies construct and evaluate untangled commits:}

\vspace{0.15cm}
\noindent 
\textbf{Limitation 1: Ignoring Runnability.}
Existing untangling methods~\cite{shen2021smartcommit,partachi2020flexeme,atomizer2026} do not check a practical and fundamental requirement for commit histories: the code should remain \emph{runnable} after each commit.
Applying a sequence of commits to a repository produces a project version after each step. We refer to the versions between the initial and final versions as \emph{intermediate versions}.
In modern development, code review and continuous integration (CI) checks are performed before changes are accepted~\cite{githubProtectedBranches}.
A commit that prevents the software from running is likely to be rejected.
For example, the Linux kernel guidelines~\cite{kernelSubmittingPatches} explicitly require the kernel to build and run properly after every patch in a series, so that developers can inspect the history to locate bugs.
Thus, we emphasize that \textit{runnability is a prerequisite of untangled histories}.
Yet existing studies~\cite{shen2021smartcommit,partachi2020flexeme,atomizer2026} do not enforce this requirement in method design and evaluation, limiting practical use.

\vspace{0.15cm}
\noindent
\textbf{Limitation 2: Unvalidated Maintenance Benefits.}
Existing studies claim that commit untangling helps developers review changes and locate bugs, but their automated evaluations do not directly test these practical benefits~\cite{partachi2020flexeme,li2022utango,atomizer2026}.
A widely used evaluation setting combines several developer-written commits from repository history into a single, synthetic tangled patch~\cite{partachi2020flexeme,li2022utango}.
Researchers then ask an untangling method to split this patch and measure how accurately it recovers the original commits.
A higher score therefore indicates closer agreement with the developers' original grouping decisions.
Atomizer's additional evaluation on expert-annotated real commits also measures agreement with the reference commits~\cite{atomizer2026}.
However, \textit{existing evaluation does not establish the practical benefits of generated commits}, especially in the era of AI coding.

These practical limitations motivate us to make two novel contributions:
\textcircled{\scriptsize 1}~\ourmethod{}, the first agentic method that untangles patches while keeping the code runnable; and
\textcircled{\scriptsize 2}~\tangleeval{}, the first evaluation framework that quantifies how untangled commits help coding agents.

\ourmethod{} (\underline{Ru}nnable \underline{C}ommit Un\underline{Tangle}) is an agentic commit untangling method.
Given a tangled patch and the initial project version, an agent groups the changes by development purpose and proposes an ordered sequence of commits.
\ourmethod{} applies these commits one by one and checks whether the code remains runnable.
When a check fails, it first uses the error message to adjust the commit order.
If failures persist, it asks the agent to revise the grouping and order.
If these revisions still fail, it merges adjacent commits to remove versions that cannot run.
It repeats the checks after each adjustment while preserving the final code produced by the original patch.
We compare \ourmethod{} with four baselines on 131 patches produced by a Claude Opus 4.6 agent across 100 Python repositories in SWE-CI~\cite{chen2026sweci}.
\ourmethod{} produces a history for every patch, with the code after every commit passing the runnability checks.
By comparison, 20.6\%--37.4\% of the histories returned by the baselines contain at least one version that fails runnability checks.

\tangleeval{} lets coding agents repair regressions in tangled, agent-generated patches.
We collect 453 patches generated by six agents from SWE-CI~\cite{chen2026sweci}, each causing previously passing tests to fail.
We ask two other coding agents, powered by \texttt{Qwen3-\allowbreak{}Coder-\allowbreak{}Next} and \texttt{Nemotron-\allowbreak{}3.5-\allowbreak{}Lightning-\allowbreak{}30B-\allowbreak{}A3B} (using mini-swe-agent~\cite{minisweagent2026} as the harness), to repair these regressions.
For each patch, the repair agents start from the same failing code.
We compare repair without untangled commits against repair with histories generated by \ourmethod{}, Armchr~\cite{armchr2026}, or Atomizer~\cite{atomizer2026}.
Because \ourmethod{} keeps the code runnable after each commit, we can run the failing tests with \texttt{git bisect} to identify a candidate commit that introduced the regression.
We add this result to the repair agent's context.
Compared with repair without untangled commits, \ourmethod{} raises pass@1 from 26.9\% to 31.3\% for Qwen and from 30.8\% to 36.0\% for Nemotron.
The average relative improvement across the two models is 16.6\%.
Armchr and Atomizer also improve repair performance, but yield smaller average relative gains of 4.5\% and 6.1\%, respectively.

To better understand how untangled commits help coding agents repair bugs, we conduct case studies comparing agent trajectories with and without these commits. We find that agents use commit diffs and messages to identify behavior that a repair should preserve. However, in one attempt, we also observe agents repeatedly reconsidering a correct repair when it conflicts with a misleading commit message.

To summarize, we make the following contributions:
\begin{itemize}
  \item \textbf{Research Gap} (Section 2):
  We identify two practical limitations of existing commit untangling research: existing commit untangling methods do not check whether the code remains runnable after each commit, and current evaluations neither consider real-world development constraints nor demonstrate the practical benefits of untangling for subsequent software development.
  
  \item \textbf{Approach} (Section 3):
  We propose \ourmethod{}, the first commit untangling method that produces runnable commits by checking runnability after each commit and uses execution feedback to revise the results.

  \item \textbf{Evaluation} (Sections 4 and 5):
  We propose \tangleeval{}, the first evaluation framework that measures how untangled commit history helps coding agents repair bugs. 

  \item \textbf{Implication} (Section 6):
  We provide the first empirical evidence that untangled histories can improve coding agents' bug-fixing performance. This finding motivates using rich histories in software development.
\end{itemize}

\section{Motivation and Background}
\label{sec:background}

\subsection{Commit Untangling and Runnability}
\label{sec:bg-untangling}

Developers use commits to record changes to a project~\cite{chacon2014progit}.
A patch is \emph{tangled} when it combines changes serving different development purposes, such as fixing a bug and performing an unrelated refactoring~\cite{herzig2013impact}.
\emph{Commit untangling} aims to separate these changes into smaller commits, each focused on one purpose~\cite{partachi2020flexeme,shen2021smartcommit,li2022utango,atomizer2026}.
This separation allows developers to inspect related changes together.
In a controlled study, Tao and Kim found that participants answered review questions more accurately when a tangled change was presented as separate groups of related changes~\cite{tao2015partitioning}.

The resulting commits should also keep the code \emph{runnable} so that developers can use the history for testing and debugging.
Applying the commits in order produces a project version after each step; the versions between the initial and final versions are the \emph{intermediate versions}.
The Linux kernel guidelines explicitly require the kernel to build and run properly after every patch in a series, because tools such as \texttt{git bisect} may inspect the code at any of these points~\cite{kernelPostingPatches}.
Runnability is distinct from functional correctness: code may run and still fail a test.
Checking only the completed patch does not establish whether the code remains runnable after each earlier commit.

\begin{figure}[t]
  \centering
  \includegraphics[width=\linewidth]{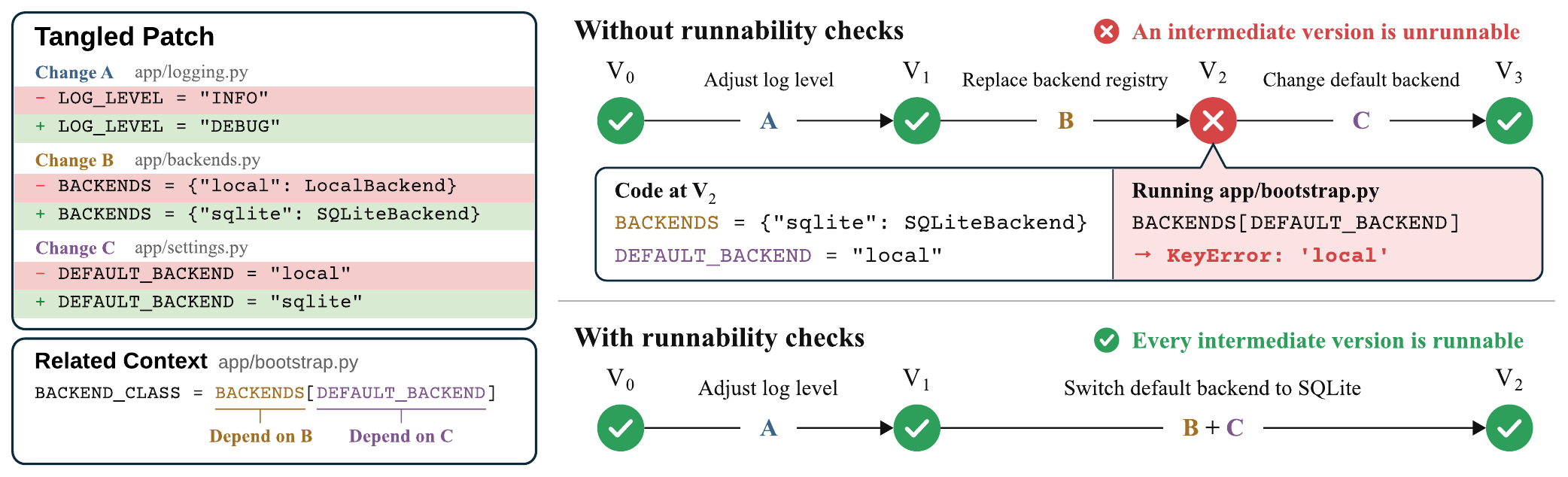}
  \caption{A simplified patch with two development purposes: adjusting the logging level (A) and switching the default backend from \texttt{local} to \texttt{sqlite} (B and C).
  Change~B replaces the entry in the backend dictionary, and Change~C updates the default selection.
  Because \texttt{bootstrap.py} evaluates \texttt{BACKENDS[DEFAULT\_BACKEND]} at import time, separating B and C causes a \texttt{KeyError} between these commits.
  Keeping A separate and grouping B with C avoids this mismatch.}
  \label{fig:utgexample}
\end{figure}

Figure~\ref{fig:utgexample} illustrates this problem with a simplified patch.
The patch adjusts the logging level through Change~A and switches the default backend through Changes~B and~C.
Change~B replaces the \texttt{local} entry in the backend dictionary with \texttt{sqlite}, while Change~C updates \texttt{DEFAULT\_BACKEND} from \texttt{local} to \texttt{sqlite}.
When \texttt{bootstrap.py} is imported, it selects the backend by evaluating \texttt{BACKENDS[DEFAULT\_BACKEND]}.
The selected name must therefore exist in the dictionary.

Suppose each change forms a separate commit.
After A and B have been applied, the dictionary contains only \texttt{sqlite}, but the default remains \texttt{local}.
Any test that imports \texttt{bootstrap.py} then fails to load with a \texttt{KeyError}.
Reversing B and C does not solve the problem: applying C first selects \texttt{sqlite} while the dictionary still contains only \texttt{local}.
Grouping B and C into one commit avoids both failures, while A can remain in a separate commit.
The runnability check exposes this mismatch, allowing the untangling method to revise its grouping.
The example highlights that runnability is a fundamental requirement in commit untangling.

\subsection{Task Formulation}
\label{sec:bg-formulation}

Let $V_0$ denote the project version before a patch $P$, and let
$V_\star = \mathrm{apply}(V_0, P)$ denote the version after applying it.
Our task is to reorganize the changes in $P$ into a commit history
$C = \langle c_1,\dots,c_k\rangle$.
Each commit contains at least one change, and every change in $P$ belongs to exactly one commit.
The commits must be applicable in sequence, producing
\[
V_i = \mathrm{apply}(V_{i-1}, c_i),
\qquad 1 \leq i \leq k.
\]
The final version must satisfy $V_k = V_\star$.
Thus, untangling changes how the patch is organized without changing its final code.
We aim to produce smaller commits organized by development purpose, while keeping together changes that cannot be separated without breaking runnability.

For the Python projects in our study, we assess runnability through test collection.
A version passes the runnability checks when \texttt{pytest} can discover and load its tests in the project environment.
Let $R(V)=1$ when version $V$ passes these checks, and $R(V)=0$ otherwise.
\emph{Runnable commit untangling} additionally requires the resulting commit history to be \emph{runnable}:
\[
R(V_i)=1,
\qquad 1 \leq i \leq k,
\]
that is, the code passes the runnability checks after every commit, including the last one.
If $V_\star$ fails these checks, preserving its code and satisfying this condition are incompatible.

We do not require functional tests to pass.
A runnable version may still contain a regression, including one introduced by the original patch.
Untangling preserves that patch; repairing its errors is a separate task.
A history containing only one commit may satisfy the requirements above, so we evaluate how well the patch is split as well as whether the resulting versions are runnable.

\subsection{Evaluating Untangled Commits}
\label{sec:bg-evaluation}

Tangled patches in practice often lack a reference showing how their changes should be grouped (i.e., the ground truth of untangled commits).
A common evaluation therefore combines several developer-written commits into one tangled patch and records which original commit each change came from~\cite{partachi2020flexeme,li2022utango}.
Methods are then scored on how closely their groups reproduce the original grouping~\cite{hou2025colauntangle,atomizer2026}.
Although the metrics differ in which code elements they count, they share this focus on agreement with reference groups.
The original commits provide a reference for comparison, but do not necessarily represent the only reasonable way to split the patch.

These grouping scores do not evaluate the runnability.
The consequences of a grouping error depend on which changes are separated.
In Figure~\ref{fig:utgexample}, separating B and C makes \texttt{bootstrap.py} fail to load after the first of these two changes, regardless of their order.
Grouping scores also do not assess commit order: rearranging the same groups leaves their scores unchanged, even if the new order imports a module before the commit that introduces it.
Finally, agreement with reference groups does not directly measure whether the history helps developers review changes or repair bugs.
These uses require evaluation beyond grouping accuracy.

\tangleeval{} evaluates both the generated histories and their use in regression repair.
For each history, it checks whether the commits can be applied in order from $V_0$, reproduce $V_\star$, and leave every resulting version runnable.
We also report commit counts to show how finely each patch is split.
For the repair evaluation, we use agent-generated patches that cause previously passing tests to fail.
For each patch, repair agents start from the same failing code and tests under the same repair budget.
We compare repair without untangled commits against repair with histories produced by different methods.
We measure how often agents fix the failing tests without causing other previously passing tests to fail.

\section{Methodology}
\label{sec:method}

This section describes how \ourmethod{} constructs commit histories that ensure runnability, and how \tangleeval{} evaluates these histories and their use in regression repair.

\subsection{\ourmethod{}: Runnable Commit Untangling}
\label{sec:method-untangling}

\begin{figure}[!t]
  \centering
  \includegraphics[width=\linewidth]{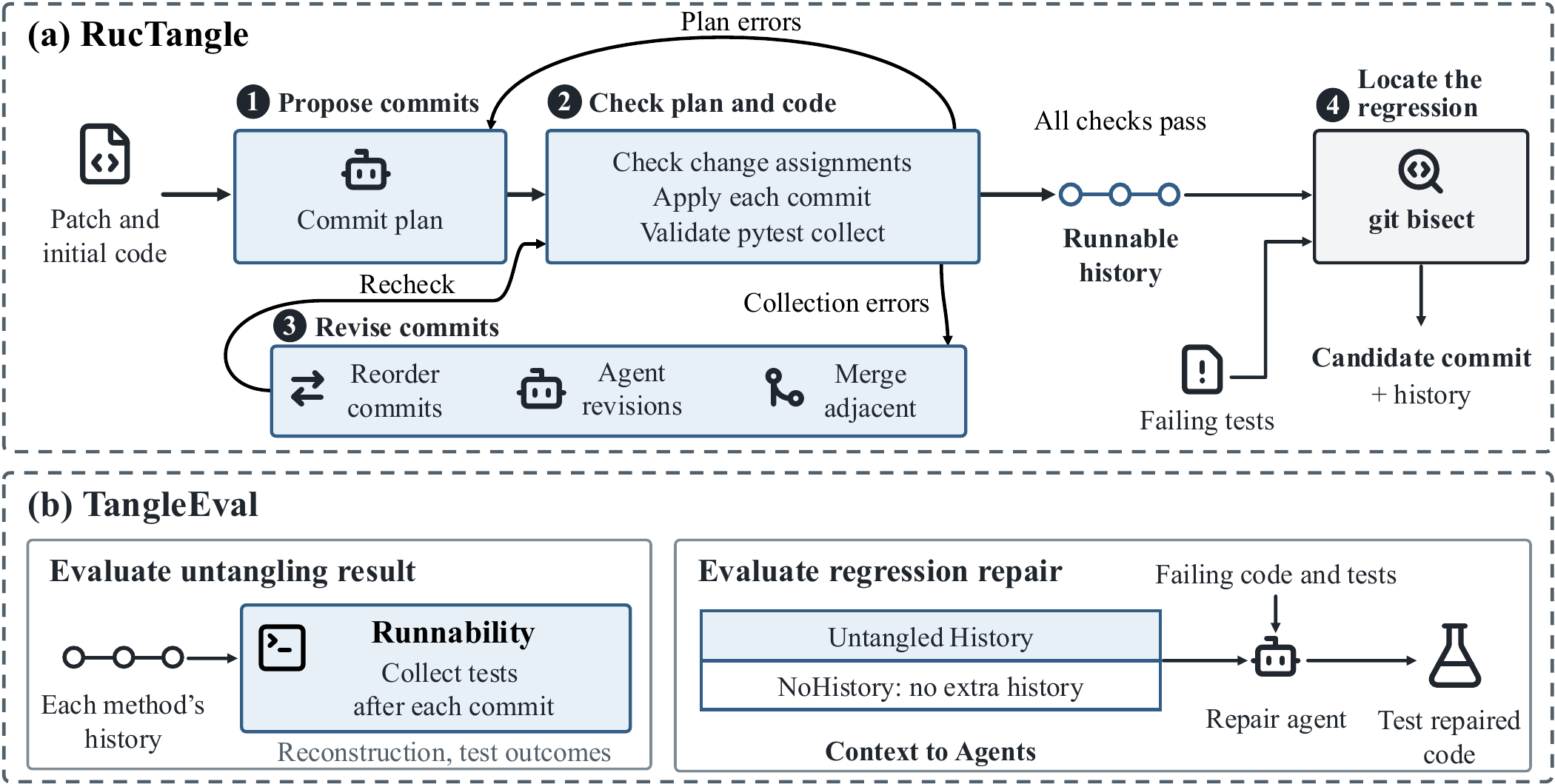}
  \caption{\ourmethod{} and its use in \tangleeval{}.
  (a) \ourmethod{} checks and revises agent-proposed ordered commits to keep each version runnable, then uses \texttt{git bisect} to identify a candidate commit for regression repair. (b) \tangleeval{} evaluates untangling results and agents' performance in regression repair with and without untangled histories.}
  \label{fig:agentuntangle-overview}
\end{figure}

Given a patch $P$ and the initial project version $V_0$, \ourmethod{} asks an agent to group the changes and propose an order for the commits.
It then applies the commits in order and checks the runnability of each resulting intermediate version.
Failures guide adjustments to the grouping and order, as shown in Figure~\ref{fig:agentuntangle-overview}.
Throughout this process, the final code must remain equal to $V_\star$.
The procedure below assumes that $V_\star$ passes the runnability checks defined in Section~\ref{sec:bg-formulation}; otherwise, preserving its code and keeping every version runnable are incompatible.

\vspace{0.1cm}
\noindent \textbf{Step 1: Proposing Commits.}
\ourmethod{} first divides $P$ into code changes for the agent to group.
Whole-file additions and deletions are initially treated as single changes.
The agent examines these changes and the project code to propose a \emph{commit plan}: an ordered list of commits, each containing a group of changes and a message describing its intended purpose.
The agent groups changes by development purpose and considers their dependencies when choosing the order.
An initial change may contain changes serving several purposes, as can happen when a newly added file contains several features.
The agent can use the \texttt{split} tool to replace such a change with smaller changes that preserve the same edits.
It then submits the completed plan through the \texttt{submit} tool.

\vspace{0.1cm}
\noindent \textbf{Step 2: Checking the Plan and Code.}
After the agent submits a commit plan, \ourmethod{} validates it in two stages.
First, it performs structural validation to ensure that the plan is well formed, that every change is assigned to exactly one commit, and that each commit contains at least one change.
Formatting or assignment errors are returned to the agent for correction and resubmission.
Next, starting from $V_0$, \ourmethod{} applies the planned commits in order to construct intermediate versions, and checks their runnability following the definition in Section~\ref{sec:bg-formulation}.

\vspace{0.1cm}
\noindent \textbf{Step 3: Revising the Commits.}
When a runnability check fails, \ourmethod{} first tries reordering the commits.
For example, an error may indicate that a required module is introduced only by a later commit; moving that commit earlier may resolve the failure without changing the commit groups.
If failures remain, \ourmethod{} starts a new agent session with the current plan and the validation error.
The agent revises the plan by regrouping or reordering changes, and further splitting changes when necessary.
\ourmethod{} allows up to two such revision sessions.
If an intermediate version $V_i$ remains unrunnable after two agent revisions, \ourmethod{} removes the corresponding commit boundary by merging $c_i$ with $c_{i+1}$.
If necessary, it progressively merges subsequent commits until the version after the merged commit becomes runnable.
This fallback removes the unrunnable intermediate version while preserving all changes, at the cost of producing fewer commits.
After each revision, \ourmethod{} revalidates the resulting plan and intermediate versions using the checks in Step~2, and accepts the plan only when all checks pass.

\vspace{0.1cm}
\noindent \textbf{Step 4: Locating the Regression.}
This step uses the runnable history to provide additional information for regression repair.
When $P$ introduces a regression, we denote $V_\star$ as $V_{\mathrm{fail}}$, and let $T_{\mathrm{fail}}$ denote the tests that pass at $V_0$ and fail at $V_{\mathrm{fail}}$.
After constructing the runnable history, \ourmethod{} runs the tests in $T_{\mathrm{fail}}$ on versions selected by \texttt{git bisect}.
The test outcomes guide the search for a candidate commit that introduced the regression.
\ourmethod{} provides the generated history and the identified candidate, including its commit message, as context for the repair agent.

\subsection{\tangleeval{}: Evaluating Untangled Commit Histories}
\label{sec:tangleeval}

\subsubsection{Evaluating Untangling Results}
\label{sec:rq1-setup}

To evaluate whether the untangling results are suitable for software development, \tangleeval{} selects large, agent-generated patches that contain changes serving multiple development purposes as the inputs for untangling, and evaluates the resulting commit histories using the following metrics:

\vspace{0.1cm}
\noindent \textbf{Patch reconstruction rate.}
A valid commit history should reconstruct the original patch: its commits can be applied sequentially from $V_0$, and the resulting final version matches $V_\star$.
This rate is the proportion of input patches for which a method produces a valid history.

\vspace{0.1cm}
\noindent \textbf{Runnability.}
\tangleeval{} checks whether the version produced after each commit passes the runnability checks defined in Section~\ref{sec:bg-formulation}.
For a history $C$ containing $k$ commits, the runnability rate $\operatorname{Run}(C)$ is the proportion of the resulting versions $V_1,\ldots,V_k$ that pass these checks.
\tangleeval{} reports the mean runnability rate across all untangling results, together with the proportion of histories for which $\operatorname{Run}(C)=1$.
The latter measures how often a method produces a history in which every intermediate version is runnable.

\vspace{0.1cm}
\noindent \textbf{Intermediate test failures.}
When commits are applied sequentially, an intermediate commit may temporarily break behavior that has already been established by preceding commits.
A desirable untangling should minimize such temporary regressions.
\tangleeval{} therefore examines whether applying a commit causes a previously passing test to fail.
Let $T$ contain the tests that pass in the final version $V_\star$.
For each untangled history, $\operatorname{ITF}(C)$ is the proportion of tests in $T$ that experience a pass-to-fail transition after applying a commit.
A lower $\operatorname{ITF}(C)$ indicates fewer temporary regressions.

\vspace{0.1cm}
\noindent \textbf{Commit count.}
The preceding metrics should be interpreted together with the degree of patch splitting, because keeping the entire patch in a single commit can reconstruct the final code without introducing any intermediate versions.
\tangleeval{} therefore reports the mean and median number of commits, together with the proportion of histories containing only one commit.
These measurements characterize how finely each method splits the input patch and serve as auxiliary metrics for interpreting the preceding results.

\vspace{0.1cm}
\noindent \textbf{Distribution of test gains.}
Commit untangling aims to generate commits that each serve a distinct development purpose.
However, directly evaluating development purposes at scale is difficult in an automated evaluation.
We therefore use the distribution of functional gains across commits as a proxy, with passing tests providing observable evidence of the corresponding behavior.
If functional gains are concentrated in only one commit, the generated history may provide limited decomposition of the patch's functionality; conversely, gains distributed across multiple commits indicate that tested behavior is established progressively across the history.
Let $T_{\mathrm{F2P}}$ contain the tests in $T$ that fail at $V_0$. These fail-to-pass (F2P) tests~\cite{jimenez2024swebench} provide evidence of functional gains between the initial and final versions.
For each F2P test, \tangleeval{} assigns the test to the commit at its last transition from failing to passing in the history.
The test passes after this commit and at every subsequent version through $V_\star$.
Let $q_i$ be the proportion of these tests assigned to commit $i$.
\tangleeval{} reports this distribution using the largest share assigned to one commit,
$\operatorname{MTP}(C)=\max_i q_i$, and its normalized entropy:
\begin{equation}
  \operatorname{TPE}(C)
  =
  -\frac{1}{\log k}
  \sum_{i:\,q_i>0}q_i\log q_i,
  \qquad k>1.
  \label{eq:tpe}
\end{equation}
We define $\operatorname{TPE}(C)=0$ when $k=1$.
A larger MTP indicates a greater concentration of test gains in one commit, while a larger TPE indicates a more even distribution across commits.

\subsubsection{Evaluating Regression Repair}
\label{sec:rq2-setup}

To evaluate whether untangled commit histories can help coding agents repair regressions, \tangleeval{} constructs repair tasks from agent-generated patches that introduce regressions. For each task, $V_0$ is the version before the patch, and $V_{\mathrm{fail}}$ is the resulting version that contains the regression. The set $T_{\mathrm{fail}}$ contains tests that pass at $V_0$ but fail at $V_{\mathrm{fail}}$, representing the regression to be repaired.
\tangleeval{} reconstructs the Git history from $V_0$ to $V_{\mathrm{fail}}$ using the commits produced by the evaluated untangling method and makes this history available to the repair agent.
For \ourmethod{}, whose reconstructed history contains runnable intermediate versions, it additionally runs the tests in $T_{\mathrm{fail}}$ on the versions selected by \texttt{git bisect} and provides the resulting bisect information as additional context to the repair agent.
The agent may then generate a \textit{repair patch}, which is applied to $V_{\mathrm{fail}}$ to obtain the repaired version $V'$.
\tangleeval{} evaluates the agent's performance by running $T_{\mathrm{fail}}$ as well as $T$ on $V'$.
Because agent behavior can vary across independent repair \textit{attempts}, \tangleeval{} supports repeated attempts.

\tangleeval{} evaluates regression repairs using three metrics. An agent achieves a \emph{partial repair} if at least one test in $T_{\mathrm{fail}}$ passes at $V'$, a \emph{full repair} if all tests in $T_{\mathrm{fail}}$ pass, and a \emph{successful repair} if, in addition, every test in $T$ passes.
\tangleeval{} reports pass@$1$, which is the proportion of attempts that achieve a \emph{successful repair}.
With multiple attempts per patch (e.g., three), \tangleeval{} also reports pass@$3$, which is the proportion of repair patches with at least one \emph{successful repair}.

To characterize repair progress beyond repair success, \tangleeval{} also reports the proportion of attempts that achieve \emph{partial repair} and \emph{full repair} to provide a more comprehensive view of the agent's performance.
Meanwhile, agents usually repair regressions through a process that involves localization, editing, and testing. \tangleeval{} records the repair patch generation rate as the proportion of attempts whose final version differs from $V_{\mathrm{fail}}$.
This rate serves as a coarse indicator of whether the agent progresses beyond regression localization to implementing a candidate repair.

\pagebreak
\section{Experimental Setup}
\label{sec:exp}

\subsection{Datasets}
\label{sec:evaluation-dataset}

\subsubsection{SWE-CI}
\label{sec:swe-ci}

SWE-CI~\cite{chen2026sweci} evaluates how coding agents extend
and maintain existing software over multiple development iterations.
Each task starts from an earlier version of a repository and uses
tests from a later version of the same project to specify the
expected behavior.
The agents must modify the earlier code until it passes these tests,
which may require adding new functionality or updating existing
behavior.
In each iteration, the benchmark runs the tests on the current code.
An architect agent uses the test failures and the code to identify
missing or incorrect behavior and write development requirements.
A programmer agent receives these requirements and modifies the
repository accordingly.
The updated code is tested again, and this cycle continues until
the tests pass or the iteration limit is reached.
We use the published release containing 226 tasks with container
images~\cite{swecidataset2026}, together with the released agent
runs~\cite{swecitrajectories2026}.
These runs record the code and test outcomes after each development
iteration.
By comparing the initial code with later recorded versions, we
extract the cumulative patches used in our evaluation.

\subsubsection{Dataset Construction}
\label{sec:dataset-construction}

We construct two datasets for different evaluation purposes.
For RQ1, we examine whether splitting a completed patch preserves
the final code and leaves runnable versions after individual commits.
For RQ2, we select patches with regressions and evaluate
whether untangled commits help another agent repair them.

Before comparing code versions, we remove a fixed set of files
created during execution, such as \texttt{pytest} caches and coverage
reports.
We retain project code, tests, configuration files, and generated
source files.
We also run tests three times
under the task's container configuration to identify consistently
passing tests.

For RQ1, we use released Claude Opus 4.6 runs whose final records
report completion of the SWE-CI task.
The input patch contains the changes from the initial version,
$V_0$, to the last recorded version, $V_\star$.
When runs share the same initial code and execution environment,
we keep only one run, giving priority to the earlier release batch.
We retain tests that pass in all three runs at $V_\star$ and require
at least one test that does not pass at $V_0$ but passes at $V_\star$.
This provides observable changes in test outcomes for evaluating
the generated commit sequence.
After filtering, the dataset contains 131 patches from 100 Python
repositories.

For RQ2, we select code versions in which an agent's changes cause previously passing tests to fail. We use runs from Claude Opus 4.6, GLM-5.1, DeepSeek-V4-Pro, Kimi-K2.6, Qwen3.7-Max, and GPT-5.5. Within each run, we track tests that pass in the initial version and record their first failing iteration and first subsequent recovery iteration. We retain failing versions with tests that pass immediately before the modification, fail after it, and later recover in the same run. After recovery, these tests have no further recorded failures and pass in the run's final version, providing evidence that the corresponding regressions are repairable.

For each retained failing version, $V_{\mathrm{fail}}$, we extract the cumulative patch from the initial version, $V_0$, to $V_{\mathrm{fail}}$ as the untangling input. The repair targets $T_{\mathrm{fail}}$ contain all tests that pass at $V_0$ but fail at $V_{\mathrm{fail}}$. We also retain the set $T$ of tests that pass at $V_{\mathrm{fail}}$ to detect new regressions introduced by a repair.
The resulting dataset contains 453 patches from 108 tasks in 90 repositories. A task can contribute multiple patches through different generating models or development iterations. Each generating model contributes between 28 patches (GPT-5.5) and 97 patches (Kimi-K2.6).

\begin{figure}[!t]
  \centering
  \includegraphics[width=0.8\linewidth]{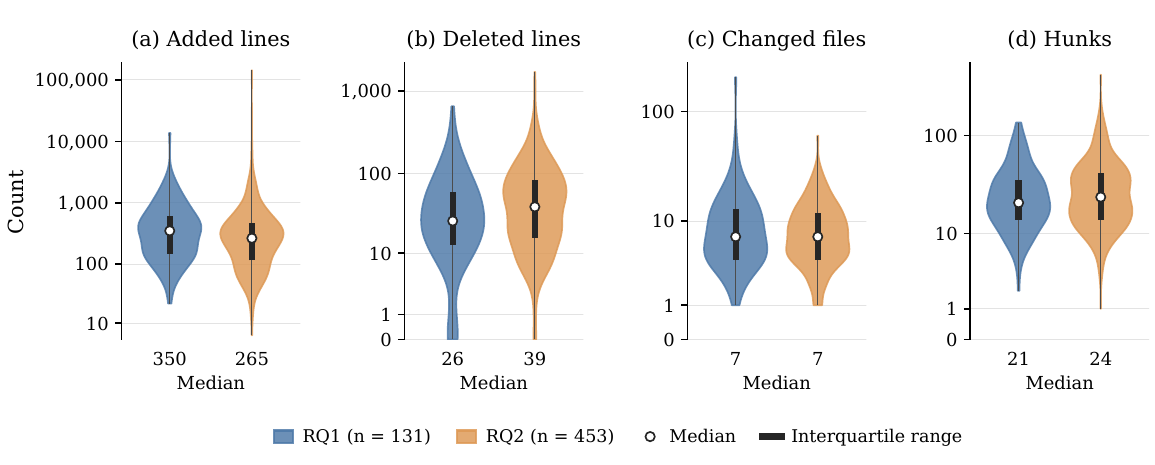}
  \caption{Distributions of modified files, diff hunks, added lines,
  and deleted lines in the RQ1 and RQ2 datasets.
  The median numbers of added and deleted lines are 350 and 26
  for RQ1, and 265 and 39 for RQ2.}
  \label{fig:patch-size}
  \label{fig:rq1-patch-size}
\end{figure}

Figure~\ref{fig:patch-size} shows the sizes of the patches in both
datasets.
Both have a median of seven modified files per patch.
The median numbers of hunks are 21 for RQ1 and 24 for RQ2.

To reduce evaluation cost, four repair models
(Section~\ref{sec:rq2-commercial-setup}) and the commit-merging
analysis (Section~\ref{sec:disc-granularity}) use the same
150-patch sample.
We draw this sample from 434 candidates for which \ourmethod{}
produces a history with at least four commits and completes
the bisect step.
The four-commit requirement ensures that merging commits to halve
their number still leaves at least two commits.
Sampling is stratified by repository and the number of commits
produced by \ourmethod{}, so that the sample approximates the
composition of the eligible candidates.
All 150 selected patches are included in the reported
453-patch repair dataset.

\subsection{Compared Methods and Repair Conditions}
\label{sec:comparison-conditions}

\subsubsection{Commit Untangling Methods}
\label{sec:untangling-methods}

We compare five commit untangling methods.
FileSplit and HunkSplit are simple structural baselines that partition a patch by file and by diff hunk, respectively.
Armchr~\cite{armchr2026} combines program-structure and dependency analysis with LLM-assisted untangling.
Atomizer~\cite{atomizer2026} constructs dependency-aware minimal change subgraphs, infers change intent, and uses multiple LLM agents to iteratively refine groups by intent.
\ourmethod{} uses an untangling agent to inspect the repository and propose an ordered commit plan, then uses runnability checks to guide revisions to the grouping and order (Section~\ref{sec:method-untangling}).

To anchor the scale of these metrics, we additionally report \textsc{NoUntangle}, a reference condition that keeps the full development patch as a single commit in the untangling results evaluation.

FileSplit, HunkSplit, and Atomizer produce unordered groups. We order these groups according to their dependencies. When no dependency determines the next group, we select the first group that can be applied to the current repository version.

We include Atomizer, a state-of-the-art commit untangling approach that outperforms prior methods on synthetic C\# and Java benchmarks~\cite{atomizer2026}.
We use a Python adaptation of Atomizer's change-graph pipeline. Our implementation uses normalized \texttt{BAAI/bge-base-en-v1.5} embeddings~\cite{xiao2024cpack} and merges groups whose cosine similarity exceeds 0.6.

\subsubsection{Regression Repair Conditions}
\label{sec:repair-conditions}

We evaluate whether providing coding agents with untangled histories helps them diagnose and repair regressions. The main comparison includes repair without untangled commits and repair with commits generated by other baseline untangling methods: Armchr~\cite{armchr2026} and Atomizer~\cite{atomizer2026}. 
FileSplit and HunkSplit are excluded because they neither group changes by development purpose nor produce commit messages.

The regression repair evaluation compares four conditions: \textsc{NoHistory}, \textsc{Armchr}, \textsc{Atomizer}, and \ourmethod{}. 
\textsc{NoHistory} provides no untangled commits. 
For the other three, \tangleeval{} provides the untangling result of each method as the context to support the repair agent.

\subsection{Implementation}

\subsubsection{Agents and Execution Environment}
\label{sec:implementation}

The untangling agent of \ourmethod{} and the repair harness are built on mini-swe-agent~\cite{minisweagent2026} (version 2.4.6). \ourmethod{}'s untangling agent, Armchr, and Atomizer all use DeepSeek-V4-Flash through the OpenRouter API. 
The two main models in regression repair evaluation, Qwen3-Coder-Next~\cite{cao2026qwen3coder} (FP8) and Nemotron-3.5-Lightning-30B-A3B~\cite{nvidia2026nemotronlightning} (BF16), are served locally with vLLM~\cite{kwon2023vllm} on four and two NVIDIA L40S 48\,GB GPUs, respectively, with a context window of 262,144 tokens.
We apply untangled commits, check runnability, and run \texttt{git bisect} and tests in the task’s SWE-CI container image, using a separate container for each version produced by an untangled commit or each repair attempt.

Each \ourmethod{} untangling session and each repair attempt is limited to 80 agent steps. \ourmethod{} uses one initial planning session and up to two revision sessions when validation fails. The repair agent may inspect and modify the repository, execute tests, and query the supplied history through Git operations such as \texttt{git log}, \texttt{git show}, \texttt{git diff}, and \texttt{git blame}. Operations that would change the checked-out revision, such as \texttt{git checkout}, are blocked. This keeps the base revision fixed at $V_{\mathrm{fail}}$ for every attempt.
All agents use temperature 0 and top-p 1. Because repair outputs can still vary across runs on the shared serving backend, the main experiment uses three repair attempts per patch and condition.

\subsubsection{Additional Models}
\label{sec:rq2-commercial-setup}

To assess whether the observed benefit extends beyond the two main models, we evaluate MiMo-V2.6-Flash, Qwen3.7-Flash, DeepSeek-V4-Flash, and GPT-6-Luna under \textsc{NoHistory} and \ourmethod{}, with one attempt per patch and condition on the shared 150-patch sample. We access these models through the OpenRouter API and run GPT-6-Luna with reasoning effort set to none. All other settings follow the main experiment.

\subsection{Trajectory Analysis}
\label{sec:rq3-setup}

Untangled histories present code changes together with commit messages describing their intended purposes. These records can help coding agents identify changes relevant to a regression and determine which behavior a repair needs to preserve. Understanding how agents interpret and act on this information can inform the design of agents that actively use software history in subsequent development. We therefore investigate how coding agents interact with untangled histories.

We conduct case studies of trajectories from the regression-repair evaluation. We select patches for which the \textsc{NoHistory} and \ourmethod{} conditions produce different repair outcomes for the same model and attempt index. For each selected patch, we compare the recorded trajectories, tracing code changes the agent inspects, how it interprets commit messages in relation to code behavior, and its subsequent repair decisions and outcomes.
When examining agents' use of commit messages, we also compare repeated attempts for the same patch, model, and condition to understand how agents interpret the same messages differently.

\section{Results Analysis}
\label{sec:results}

\subsection{RQ1. How well do different untangling methods split AI-generated patches?}
\label{sec:rq1-results}

We evaluate methods on 131 agent-generated patches.
Table~\ref{tab:rq1-quality} reports whether their outputs reconstruct the original patches, remain runnable after each commit, and introduce temporary test failures.
It also describes the number of commits and the distribution of test gains.
\textsc{NoUntangle} keeps each complete patch in a single commit and serves as a reference for interpreting results.

\begin{table}[!t]
  \centering
  \caption{Untangling results on 131 patches. Metric definitions are given in Section~\ref{sec:rq1-setup}.
  Bold marks the best values among the five untangling methods, except for the commit count metrics. Lower values are better for mean Intermediate test failures (ITF) and MTP.
  \textsc{NoUntangle} is a reference condition.}
  \label{tab:rq1-quality}
  \footnotesize
  \setlength{\tabcolsep}{2.2pt}
  \begin{tabular}{@{}lrrrrrrrrrr@{}}
    \toprule
    & & \multicolumn{3}{c}{Commit count}
    & \multicolumn{2}{c}{Runnability}
    & \multicolumn{2}{c}{ITF}
    & \multicolumn{2}{c}{Test gains} \\
    \cmidrule(lr){3-5}
    \cmidrule(lr){6-7}
    \cmidrule(lr){8-9}
    \cmidrule(lr){10-11}
    Method
    & \shortstack{Patch\\reconstruction}
    & Mean $k$
    & Med.\ $k$
    & \shortstack{Single commit\\(\%)}
    & \shortstack{Mean\\rate}
    & \shortstack{All versions\\(\%)}
    & \shortstack{Mean\\rate}
    & \shortstack{No failures\\(\%)}
    & MTP
    & TPE \\
    \midrule
    \textsc{NoUntangle}
      & 131 (100\%) & 1.00 & 1 & 100.0
      & 1.000 & 100.0 & 0.000 & 100.0 & 1.000 & 0.000 \\
    \midrule
    FileSplit
      & \textbf{131 (100\%)} & 11.18 & 7 & 2.3
      & 0.866 & 72.5 & 0.209 & 57.7 & 0.693 & 0.400 \\
    HunkSplit
      & \textbf{131 (100\%)} & 29.71 & 21 & 0.0
      & 0.830 & 62.6 & 0.324 & 30.2 & 0.610 & 0.361 \\
    Armchr
      & 126 (96.2\%) & 6.54 & 5 & 7.1
      & 0.898 & 79.4 & 0.164 & 59.2 & 0.739 & 0.394 \\
    Atomizer
      & 126 (96.2\%) & 7.52 & 6 & 3.2
      & 0.912 & 73.8 & 0.207 & 51.2 & 0.699 & 0.419 \\
    \ourmethod{}
      & \textbf{131 (100\%)} & 8.64 & 8 & 0.8
      & \textbf{1.000} & \textbf{100.0}
      & \textbf{0.017} & \textbf{82.3} & \textbf{0.552} & \textbf{0.562} \\
    \bottomrule
  \end{tabular}
  \par\smallskip
\end{table}

We observe that
\textit{reconstructing the original patch does not ensure that the code remains runnable after each commit.}
All five methods reconstruct at least 96.2\% of the input patches.
However, only 62.6\%--79.4\% of the reconstructed histories from the four baselines pass the runnability checks after every commit.
For example, \textsc{FileSplit} and \textsc{HunkSplit} reconstruct all 131 patches, but their proportions of fully runnable histories are 72.5\% and 62.6\%, respectively.
\ourmethod{} meets both requirements on all 131 patches.
These results show that preserving the final code does not by itself make the generated history suitable for execution at each commit.

\ourmethod{} achieves full runnability while splitting nearly all patches into multiple commits.
\textsc{NoUntangle} achieves full runnability and no intermediate test failures by retaining each patch as one commit.
This reference shows why execution results must be interpreted alongside the amount of splitting.
\ourmethod{} produces a median of eight commits per patch, and only 0.8\% of its histories contain a single commit.
Armchr and Atomizer produce medians of five and six commits, with single-commit histories accounting for 7.1\% and 3.2\% of their outputs, respectively.
\textsc{FileSplit} creates one commit per changed file, while \textsc{HunkSplit} creates one per diff hunk.
They produce more commits on average than \ourmethod{}: 11.18 and 29.71, respectively, compared with 8.64.
However, they have the lowest proportions of histories runnable after every commit.

A smaller proportion of tests temporarily fail in histories generated by \ourmethod{}.
Among the tests that pass in the final version, we measure the fraction that change from passing to failing after at least one commit.
Averaged across the evaluated histories, this fraction is 1.7\% for \ourmethod{}, compared with 16.4\%--32.4\% for the four baselines.
Thus, fewer such tests temporarily fail in \ourmethod{}'s histories.
This result is observed even though \ourmethod{} does not use the tests' pass/fail results to revise the commit grouping or order.

Test gains are less concentrated in histories generated by \ourmethod{}, as measured by the metrics in Section~\ref{sec:rq1-setup}.
Its mean MTP is 0.552, compared with 0.610--0.739 for the baselines, and its mean TPE is 0.562, compared with 0.361--0.419.
The comparison with \textsc{HunkSplit} shows why commit counts alone do not describe functional decomposition: \textsc{HunkSplit} produces more than three times as many commits on average, but has a higher mean MTP of 0.610.
Finer structural splitting therefore does not necessarily reduce the concentration of test gains.
These observations suggest that, compared to other untangling methods, \ourmethod{} can distribute a tangled patch's functional gains more evenly across commits, indicating that \ourmethod{}'s histories are more likely to provide commits that serve single development purposes.

Regarding the cost, \ourmethod{} has the lowest average untangling time among the three LLM-based methods. Across all 131 patches, including unsuccessful attempts, \ourmethod{} takes 276.3 seconds per patch on average, 74.6\% of Armchr's time and 18.0\% of Atomizer's.
It generates an average of 11.9K output tokens per patch, close to Armchr's 11.7K and only 25.9\% of Atomizer's 46.0K.
These measurements include constructing, checking, and revising the history, but exclude the subsequent bisect step and regression repair.

\begin{keybox}{Answer to RQ1}
\ourmethod{} is the only tested untangling method to reconstruct all 131 patches while keeping the code runnable after every commit, with a median of eight commits per patch.
It also temporarily breaks a smaller proportion of tests: among tests that pass in the final version, an average of 1.7\% change from passing to failing during the commit sequence, compared with 16.4\%--32.4\% for the baselines.
Its histories show less concentrated test gains across commits than baseline histories.
\end{keybox}

\subsection{RQ2. How does commit untangling affect regression repair by coding agents?}
\label{sec:rq2-results}

For each patch, repair agents start with the same failing code, task description and failure information, and the same repair budget.
Under \textsc{NoHistory}, they receive no untangled commits.
We provide the information generated by \ourmethod{}, \textsc{Armchr} and \textsc{Atomizer} to the agent context.
In total, we evaluate six models.
\texttt{Qwen3-Coder-Next} and
\texttt{Nemotron} are evaluated on 453 patches, with three attempts per patch under each condition.
MiMo-V2.6-Flash, Qwen3.7-Flash, DeepSeek-V4-Flash, and GPT-6-Luna are evaluated under \textsc{NoHistory} and \ourmethod{} on the shared 150-patch sample (Section~\ref{sec:dataset-construction}), with one attempt per patch due to budget limits.

A successful repair makes all target tests pass without breaking other evaluation tests that passed before repair.
Pass@$1$ measures the proportion of successful attempts.
For the models with three attempts per patch, pass@$3$ measures the proportion of patches successfully repaired in at least one attempt.
Partial and full repair indicate whether at least one or all target tests pass, respectively; both can include new test failures.

\begin{table}[!t]
  \centering
  \caption{Repair results for Qwen3-Coder-Next and Nemotron on 453 patches, with three attempts per patch and condition.
  The baseline \textsc{Armchr} and \textsc{Atomizer} conditions provide commits from corresponding untangling methods.
  The \ourmethod{} condition provides its commits together with bisect results.
  When the method fails, we reuse the corresponding \textsc{NoHistory} outcomes.
  All values are percentages, with metric definitions given in Section~\ref{sec:rq2-setup}.
  Bold and underlined values mark column maxima within each model.}
  \label{tab:rq2-main}
  \small
  \setlength{\tabcolsep}{5pt}
  \begin{tabular}{@{}llrrrrr@{}}
    \toprule
    \multicolumn{1}{@{}r}{Model}
      & \multicolumn{1}{r}{Condition}
      & \multicolumn{1}{r}{pass@$1$}
      & \multicolumn{1}{r}{pass@$3$}
      & \multicolumn{1}{r}{Partial repair}
      & \multicolumn{1}{r}{Full repair}
      & \multicolumn{1}{r@{}}{Patch generation} \\
    \midrule
    Qwen3-Coder-Next & \textsc{NoHistory}
      & 26.9\%
      & 40.4\%
      & 43.9\%
      & 32.8\%
      & 56.0\% \\
    & \textsc{Armchr}
      & 28.8\%
      & 40.4\%
      & 47.2\%
      & 36.1\%
      & 58.6\% \\
    & \textsc{Atomizer}
      & 28.8\%
      & 40.4\%
      & 47.6\%
      & 35.7\%
      & 57.6\% \\
    & \ourmethod{}
      & \underline{\textbf{31.3\%}}
      & \underline{\textbf{43.5\%}}
      & \underline{\textbf{51.1\%}}
      & \underline{\textbf{39.7\%}}
      & \underline{\textbf{63.0\%}} \\
    \midrule
    Nemotron & \textsc{NoHistory}
      & 30.8\%
      & 41.1\%
      & 61.0\%
      & 41.2\%
      & 80.4\% \\
    & \textsc{Armchr}
      & 31.3\%
      & 43.7\%
      & 62.5\%
      & 42.2\%
      & 83.8\% \\
    & \textsc{Atomizer}
      & 32.3\%
      & 44.2\%
      & 62.5\%
      & 42.9\%
      & 82.4\% \\
    & \ourmethod{}
      & \underline{\textbf{36.0\%}}
      & \underline{\textbf{47.7\%}}
      & \underline{\textbf{69.5\%}}
      & \underline{\textbf{49.4\%}}
      & \underline{\textbf{88.3\%}} \\
    \bottomrule
  \end{tabular}
\end{table}

Our experiment shows that \textit{\ourmethod{} improves repair success for four of the six evaluated models.}
Compared with \textsc{NoHistory}, \ourmethod{} raises Qwen3-Coder-Next's pass@$1$ from 26.9\% to 31.3\% and Nemotron's from 30.8\% to 36.0\%.
Armchr and Atomizer also help these two models repair more patches, but achieve smaller improvements than \ourmethod{}.
For MiMo-V2.6-Flash and Qwen3.7-Flash, \ourmethod{} raises pass@$1$ from 29.3\% to 36.0\% and from 50.0\% to 53.3\%, respectively.
However, \ourmethod{} does not improve DeepSeek-V4-Flash's pass@$1$, which remains at 61.3\%.
GPT-6-Luna has a lower pass@$1$ with \ourmethod{}, 48.7\% against 50.0\% under \textsc{NoHistory}, corresponding to 73 successful repairs instead of 75 on the 150-patch sample.

For Qwen3-Coder-Next and Nemotron, \ourmethod{} also achieves the highest pass@$3$.
Compared with \textsc{NoHistory}, the rates increase from 40.4\% to 43.5\% and from 41.1\% to 47.7\%, respectively.
Thus, more patches receive at least one successful repair within three attempts.
For Qwen3-Coder-Next, Armchr and Atomizer increase pass@$1$ but leave pass@$3$ unchanged at 40.4\%.

Additionally, we observe that \textit{agents can repair the target regressions while introducing new ones.}
With \ourmethod{}, all six models more often fix the regressions targeted by the repair task.
However, five models also more often cause previously passing tests to fail, indicating that their repairs introduce new regressions.
On the 150-patch sample, \ourmethod{} raises GPT-6-Luna's full repair rate from 70.7\% to 79.3\%.
However, its rate of introducing new regressions also increases from 34.0\% to 43.3\%.
Its pass@$1$ falls from 50.0\% to 48.7\%, since this metric also requires previously passing tests to remain passing.
For DeepSeek, \ourmethod{} similarly raises the full repair rate but leaves pass@$1$ unchanged.
For Qwen3-Coder-Next and Nemotron, \ourmethod{} improves pass@$1$, although their rates of introducing new regressions also increase.

Patch generation also increases for all six models.
This indicates that more attempts leave code changes, including changes left when an agent reaches its step limit.
The repair metrics establish whether those changes fix the original failures and preserve previously passing tests.
These results show why checking only the original failures would overstate repair success.
Repairs made with untangled histories should also be checked against tests that passed before the repair.
Section~\ref{sec:rq3-results} examines individual trajectories to understand how agents use the supplied information when choosing and implementing repairs.

\begin{keybox}{Answer to RQ2}
\ourmethod{} improves pass@$1$ for four of the six evaluated models, leaves one unchanged, and yields a slightly lower rate for one.
It also achieves the highest pass@$3$ for both models evaluated with repeated attempts.
Across all six models, the original failures are fixed more often, but new test failures also become more frequent for most models, limiting the improvement in overall repair success.
\end{keybox}

\subsection{RQ3. How do coding agents utilize untangled histories?}
\label{sec:rq3-results}

We examine some repair trajectories from RQ2 to understand how agents use untangled histories.
The following three observations describe how agents inspect changes, interpret commit messages, and decide what to repair.
We illustrate each observation with concrete attempts on the same patch.

\begin{table}[!t]
  \centering
  \caption{Repair outcomes of MiMo-V2.6-Flash, Qwen3.7-Flash, DeepSeek-V4-Flash, and GPT-6-Luna on the shared 150-patch sample, with one attempt per patch and condition.
  All outcome columns are percentages with denominator 150.
  Partial repair counts attempts that fix at least one test failure, including full repairs.
  Bold and underlined values mark column maxima within each model.}
  \label{tab:rq2-commercial}
  \small
  \setlength{\tabcolsep}{5pt}
  \begin{tabular}{@{}llrrrr@{}}
    \toprule
    \multicolumn{1}{r}{Model}
      & \multicolumn{1}{r}{Condition}
      & \multicolumn{1}{r}{pass@$1$}
      & \multicolumn{1}{r}{Partial repair}
      & \multicolumn{1}{r}{Full repair}
      & \multicolumn{1}{r@{}}{Patch generation} \\
    \midrule
    MiMo-V2.6-Flash & \textsc{NoHistory}
      & 29.3\%
      & 38.0\%
      & 35.3\%
      & 42.0\% \\
    & \ourmethod{}
      & \underline{\textbf{36.0\%}}
      & \underline{\textbf{45.3\%}}
      & \underline{\textbf{41.3\%}}
      & \underline{\textbf{51.3\%}} \\
    \midrule
    Qwen3.7-Flash & \textsc{NoHistory}
      & 50.0\%
      & 81.3\%
      & 62.7\%
      & 88.7\% \\
    & \ourmethod{}
      & \underline{\textbf{53.3\%}}
      & \underline{\textbf{88.7\%}}
      & \underline{\textbf{73.3\%}}
      & \underline{\textbf{95.3\%}} \\
    \midrule
    DeepSeek-V4-Flash & \textsc{NoHistory}
      & \underline{\textbf{61.3\%}}
      & 86.0\%
      & 72.0\%
      & 91.3\% \\
    & \ourmethod{}
      & \underline{\textbf{61.3\%}}
      & \underline{\textbf{92.0\%}}
      & \underline{\textbf{79.3\%}}
      & \underline{\textbf{93.3\%}} \\
    \midrule
    GPT-6-Luna & \textsc{NoHistory}
      & \underline{\textbf{50.0\%}}
      & 87.3\%
      & 70.7\%
      & 96.0\% \\
    & \ourmethod{}
      & 48.7\%
      & \underline{\textbf{94.7\%}}
      & \underline{\textbf{79.3\%}}
      & \underline{\textbf{98.0\%}} \\
    \bottomrule
  \end{tabular}
\end{table}

\vspace{0.1cm}
\noindent \textit{Observation 1: Agents use history to check what changed before deciding what to repair.}
Several parts of the code may affect the same failing behavior.
Commit diffs and earlier versions let an agent check which parts were changed by the patch and which were already present.

In one task from our dataset, a coding agent modified the
\texttt{poethepoet} project.
It mistakenly deleted code that adds the required line breaks
at the end of the program's command-line help text.
The resulting help text no longer matched the expected format,
causing eight previously passing tests to fail.
In its first \textsc{NoHistory} attempt, \texttt{Nemotron} removes the helper's \texttt{rstrip()} and another trimming call without restoring the deleted newline handling.
The resulting patch fixes none of the eight target test failures.
In its first \ourmethod{} attempt, \texttt{Nemotron} inspects the candidate commit supplied by \ourmethod{}'s \texttt{git bisect} step and restores the deleted code.
The third attempt shows another use of history: the agent also suspects \texttt{rstrip()}, but checks an earlier version and confirms that the call was already present.
It leaves the helper unchanged and restores the newline handling.
All three \ourmethod{} attempts fix all eight target failures without introducing additional test failures, whereas none of the three \textsc{NoHistory} attempts achieves a full repair.

\vspace{0.1cm}
\noindent \textit{Observation 2: Agents consult commit messages to check which behavior a repair should preserve.}
A change introducing errors may also add behavior that should be retained.
By reading its commit message, agents check why the change was made and decide what to preserve while repairing.

In another example, a coding agent introduced an error while modifying XML parsing in the \texttt{pydantic-xml} project.
It added a \texttt{consume()} call to mark an XML element as processed, preventing the parser from reporting it again as unexpected input.
However, this call also clears the element's contents.
The agent placed it before the code that reads those contents for the result, causing a previously passing test to fail.
In one \textsc{NoHistory} repair attempt, \texttt{Qwen} recognizes the call's purpose from the code documentation but ultimately removes it.
The original failing test then passes, but three previously passing tests fail because processed elements are now reported as unexpected input.
In one \ourmethod{} repair attempt, \texttt{Qwen} consults the commit message to check why \texttt{consume()} was added.
The message explains that processed elements should no longer be treated as unhandled input.
The agent then examines the code that checks for unexpected input and retains the call.
It changes the order so that the parser reads and saves the contents before calling \texttt{consume()}.
This preserves the returned data while still marking the element as processed.
The resulting patch fixes the original test failure without causing additional test failures.

\vspace{0.1cm}
\noindent \textit{Observation 3: Agents may be misled by commit messages.}
A generated commit message may incorrectly describe a faulty change as an improvement.
An agent may then question a correct repair because it conflicts with the explanation in the message.

In the \texttt{poethepoet} case described above, \ourmethod{} locates the commit that deleted the required newline handling.
However, the generated message calls these line breaks ``redundant'' because ``callers already handle line breaks.''
The diff shows the deletion responsible for the failing output, but the message gives a reason to keep that deletion.
In another \ourmethod{} case, \texttt{Qwen} identifies the deleted code and repeatedly proposes restoring it.
However, it keeps returning to the message and questions whether the tests should change instead.
Along with repeatedly checking the number of line breaks, it keeps revisiting this uncertainty until reaching the step limit without changing the code.
In the second attempt, \texttt{Qwen} receives the same commit and message and raises similar doubts, but eventually restores the deleted code and tests the repair, fixing all target test failures without introducing any additional failures.

This observation motivates further study of how commit messages generated during untangling affect agents' repair decisions.
Future work could adapt commit message generation methods~\cite{dong2022fira,li2024omg} to describe and explain untangled changes more accurately, helping agents understand development purposes~\cite{tian2022goodcommit,li2023commitmatters}.

\begin{keybox}{Answer to RQ3}
Agents use commit diffs and earlier code to check which changes need repair.
They consult commit messages to understand why a change was made and decide what behavior to preserve.
However, misleading messages may lead agents to question a correct repair, even after the faulty change has been located.
\end{keybox}

\section{Discussion}
\label{sec:discussion}

\subsection{How Do Coarser Histories Affect Repair?}
\label{sec:disc-granularity}

\ourmethod{} does not prescribe an exact number of untangled commits.
The agent chooses how to group the changes, subject to a limit of 20 commits, and subsequent revisions may change this grouping.
Separating dependent changes can leave the code unable to run before all required changes have been applied.
Nevertheless, in RQ1, \ourmethod{} keeps the code runnable after every commit while producing a median of 8 commits per patch, compared with 5 for Armchr and 6 for Atomizer.
We now examine whether this finer-grained decomposition contributes to regression repair, or whether commits can be merged without reducing repair success.

We use the shared sample of 150 patches described in Section~\ref{sec:dataset-construction}.
Each selected patch has a \ourmethod{} history containing at least four commits. We compare two settings: (1) \ourmethod{}-\textsc{Original}, which retains the history generated by \ourmethod{}, and (2) \ourmethod{}-\textsc{Halved}, which reduces a history of $k$ commits to $\lceil k/2\rceil$ commits.
To construct \ourmethod{}-\textsc{Halved}, we randomly retain $\lceil k/2\rceil-1$ of the $k-1$ boundaries between adjacent commits and merge across the remaining boundaries.
This produces consecutive groups that preserve the original order but need not contain the same number of commits.
The commit messages are concatenated if the commits are merged.

Merging removes intermediate versions but does not modify the code at the retained versions. The initial and final versions therefore remain unchanged, and every retained intermediate version is drawn from the original runnable history. Consequently, \ourmethod{}-\textsc{Halved} remains runnable after every commit. We rerun the bisect step on the merged history and provide the resulting candidate commit and history to the repair agent. All other repair settings follow the \ourmethod{} condition in RQ2. To limit evaluation cost, we evaluate only \texttt{Nemotron}, using three repair attempts per patch and a limit of 80 agent steps per attempt.
Table~\ref{tab:disc-granularity} compares the two conditions.

\begin{table}[!t]
  \centering
  \caption{Nemotron repair outcomes with original and halved histories on 150 patches, with three attempts per patch.
  Both settings provide the history and its bisect result.
  Bold and underlined values mark column maxima, except for new regression, where the minimum is marked.}
  \label{tab:disc-granularity}
  \small
  \setlength{\tabcolsep}{5pt}
  \begin{tabular}{@{}lrrrrr@{}}
    \toprule
    \multicolumn{1}{@{}r}{History}
      & \multicolumn{1}{r}{pass@$1$}
      & \multicolumn{1}{r}{Partial repair}
      & \multicolumn{1}{r}{Full repair}
      & \multicolumn{1}{r}{New regression}
      & \multicolumn{1}{r@{}}{Patch generation} \\
    \midrule
    \ourmethod{}-\textsc{Original}
      & \underline{\textbf{34.9\%}}
      & \underline{\textbf{65.1\%}}
      & \underline{\textbf{46.9\%}}
      & \underline{\textbf{25.6\%}}
      & 83.3\% \\
    \ourmethod{}-\textsc{Halved}
      & 31.8\%
      & 64.7\%
      & 45.1\%
      & 28.4\%
      & \underline{\textbf{86.4\%}} \\
    \bottomrule
  \end{tabular}
\end{table}

We observe that pass@$1$ decreases from 34.9\% with \ourmethod{}-\textsc{Original} to 31.8\% with \ourmethod{}-\textsc{Halved}.
To better understand this difference, we examine the other repair outcomes.
Partial repair counts attempts that fix at least one target failure, while full repair requires all target failures to be fixed.
The partial repair rate changes little, from 65.1\% to 64.7\%, but the full repair rate decreases from 46.9\% to 45.1\%.
New regressions occur when a repair breaks tests that previously passed; their rate increases from 25.6\% to 28.4\%.
Thus, after merging, fewer attempts fix all target failures and more introduce failures elsewhere.

We also examine patch generation, which counts attempts that leave a nonempty diff, including edits left by attempts that reach their limit.
This rate increases from 83.3\% to 86.4\%.
The agent therefore leaves changes more often with \ourmethod{}-\textsc{Halved}, but fewer attempts satisfy the complete repair requirements.
These observations suggest that retaining the original histories may help the agent complete repairs while preserving existing behavior.
However, merging also changes the scope of the candidate commit identified by bisect and the messages presented together.
The results reflect this combined change in the information available to the agent, rather than the effect of commit count alone.

\subsection{Implications and Research Opportunities}
\label{sec:implications}

\paragraph{Adopting established software engineering practices.}
Our results illustrate the continued value of established software engineering practices in agentic coding.
Organizing changes into runnable histories and providing bisect results improves repair for several models without retraining them.
Recent training pipelines, such as SWE-Gym and SWE-smith, use test outcomes to identify successful coding trajectories~\cite{pan2025swegym,yang2025swesmith}.
Passing these tests alone does not establish whether the resulting changes are easy to inspect and maintain; their usability must also be considered~\cite{yang2024properties}.
Agent training and development workflows should therefore include preparing changes for later review and maintenance, alongside completing the current coding task.

\paragraph{Training agents to use project history.}
Providing history does not ensure that an agent will use it effectively.
This motivates research on \emph{history-aware coding agents}: agents trained to identify relevant commits, compare earlier and current code, and check historical explanations against code and tests.
Training tasks could include both useful and misleading historical information, allowing researchers to examine whether agents learn when to rely on it.
Evaluation should measure whether this training improves repair decisions while preserving previously working behavior.

\paragraph{Designing project histories for subsequent agents.}
Alongside commits and review discussions, recorded agent runs can preserve which files were inspected, which edits were attempted, what tests returned, and how the agent explained its decisions.
These records could help later agents understand a change, but repeated exploration, abandoned approaches, and mistaken explanations may obscure relevant information.
Future systems could link selected actions and test results to commits and let agents retrieve further details as needed to assess changes without reconstructing the entire development process.
Recorded explanations should also remain distinguishable from observed edits and test outcomes.

\paragraph{Evaluating commit messages through subsequent decisions.}
We did not independently control message quality in our experiments, leaving its effect on repair as an open research question.
Beyond accuracy and readability, evaluation could examine whether a message helps an agent identify relevant changes, preserve required behavior, and avoid introducing new regressions.

\subsection{Threats to Validity}
\label{sec:disc-threats}

\vspace{0.1cm}
\noindent\textbf{Threats to Internal Validity.}
Our comparison with Atomizer depends on its Python adaptation, including the embedding model, grouping threshold, and commit ordering.
We document these choices, use the same LLM across the three LLM-based untangling methods, and also compare with Armchr's independent implementation.
We use the project containers and select tests with consistent outcomes across three executions of candidate snapshots.
The main repair experiment uses three attempts per patch and condition, with the same failing code, tests, and repair budget across conditions.
The additional-model experiment uses one attempt per patch and provides less evidence about variation across runs.
Our repair comparisons evaluate the complete \ourmethod{} system, including bisect localization, without isolating component contributions.

\vspace{0.1cm}
\noindent\textbf{Threats to External Validity.}
Our experiments cover multiple repositories and models, but are limited to Python projects from SWE-CI.
The additional-model results show that the repair benefit varies across models, and the merging analysis concerns only histories with at least four commits and available localization results.
We focus on agent-based regression repair because coding agents increasingly participate in maintenance, and tests allow us to compare repair outcomes automatically.
Whether untangled histories also help agents perform code review and refactoring, or help human developers review and integrate changes, remains to be evaluated.
The checking stage's effectiveness for projects in other languages or with different deployment requirements remains to be evaluated.

\section{Related Work}
\label{sec:related}

\subsection{Commit Untangling}
\label{sec:related-untangling}

Commit untangling separates changes that serve different development purposes.
Early studies show that tangled changes obscure the relationship between code changes and bug reports and affect defect prediction~\cite{herzig2013impact,herzig2016impact}.
Methods therefore seek to identify which changes belong together.
EpiceaUntangler~\cite{dias2015untangling} groups recorded edits using time gaps, intervening edit counts, and whether they modify the same class.
Flexeme~\cite{partachi2020flexeme} groups changes using code dependencies and the identifiers involved in those dependencies.
UTango~\cite{li2022utango} learns to group changes using their dependencies and surrounding code.
Recent methods use LLMs to interpret the purposes of changes.
CoLaUntangle~\cite{hou2025colauntangle} combines agents that examine code dependencies and conceptual relationships, then review and revise their proposed groups.
Atomizer~\cite{atomizer2026} asks agents to infer each change's purpose and check whether grouped changes serve the same purpose.

Dependencies also matter when groups of changes become separate commits.
SmartCommit~\cite{shen2021smartcommit} constrains grouping to avoid compilation errors and orders commits so that required changes appear before their dependents.
Armchr~\cite{armchr2026}, which we include in our experiments, also groups changes and orders the resulting patches using their dependencies.
Earlier tools address related tasks: Crisp~\cite{ren2006crisp} constructs compilable versions containing selected changes to help developers locate failures, while ChgCutter~\cite{guo2019decomposing} helps developers select related changes and the tests affected by them.
Safe-commit~\cite{wloka2009safecommit} identifies subsets of workspace changes that can be committed without causing additional failures of existing tests.
These tools establish the importance of dependencies, but still ignore the runnability requirement that this paper emphasizes.

\ourmethod{} constructs an ordered history that preserves every change in the input patch.
It applies the proposed commits and checks whether the tests can be collected after each commit.
Our experiments compare this approach with Armchr and a Python adaptation of Atomizer, alongside file-based and hunk-based splitting. Our experiments exclude other works~\cite{partachi2020flexeme,li2022utango,shen2021smartcommit} designed for specific programming languages.

\subsection{Evaluating Untangled Commits}
\label{sec:related-safety}

A common evaluation combines several developer-written commits into one patch and asks an untangling method to recover the original groups~\cite{partachi2020flexeme,li2022utango}.
Atomizer~\cite{atomizer2026} also compares its output with expert annotations of real commits.
These evaluations measure agreement with a reference grouping, which need not be the only reasonable way to split a patch.
Developer studies provide another form of evidence: Barnett et al.~\cite{barnett2015decomposition} investigate whether developers find automatically separated changes useful for understanding code reviews.
Recent benchmarks evaluate histories produced by coding agents.
GitGoodBench~\cite{lindenbauer2025gitgoodbench} includes a task in which agents organize uncommitted changes into commits and an LLM assesses the resulting history.
Its iterative committing setup restricts the agent to changes within one file.
AtomicCommitBench~\cite{lin2026atomiccommitbench} evaluates histories reconstructed from squashed developer commits.
It measures agreement with the original groups and checks whether the generated commits can be applied in order.
It also separately reverses each generated commit's changes from the final code and reruns selected tests to examine whether the resulting failures are confined to particular commits.

\tangleeval{} examines the versions produced as commits are applied in order.
For agent-generated patches without a reference grouping, it checks whether the history reconstructs the original patch, whether the code remains runnable after each commit, and whether tests temporarily fail along the history.
It also evaluates the use of these histories in regression repair.
This extends the evaluation from how commits are grouped and how their changes affect tests to whether coding agents can use the generated histories to repair errors.

\section{Conclusion}
\label{sec:conclusion}

We presented \ourmethod{}, an agentic method that uses execution feedback to regroup changes and reorder commits while preserving the final code of the original patch.
On 131 agent-generated patches, \ourmethod{} produces histories that pass the runnability checks after every commit, with a median of eight commits per patch.
We also introduced \tangleeval{} to evaluate how untangled histories help coding agents repair regressions.
On 453 patches that introduce regressions, providing \ourmethod{}'s histories together with \texttt{git bisect} results raises pass@1 by 16.6\% on average across two models, relative to repair without untangled commits.
Our case studies show how agents use commit diffs to identify relevant changes and consult commit messages to preserve required behavior.
They also reveal an attempt in which an agent repeatedly questions a correct repair after reading a misleading message and exhausts its budget without producing a patch.

These results demonstrate how an established software engineering practice can support coding agents in maintaining code.
They motivate further research on how agents can actively use software history, while checking its explanations against code and tests, to guide development decisions.

\bibliographystyle{assets/plainnat}
\bibliography{references}

\end{document}